# Delayed DNS: Crippling Crypto-Ransomware

JONATHAN GRAHAM, OCAD University

This research seeks to expose a major weakness in Crypto-ransomware by modeling it as four integral sub-systems consisting of: An Agent, a Command and Control Service (CNC), an anonymous payment channel (APC) and an obfuscated command channel (OCC). We will show that most modern counter-measures focus on either the Agent or the CNC subsystems and usually in a reactive way exposing the target to undue risk. However, by disrupting this fourth component – the Obfuscated Command Channel – we can proactively and safely defeat a wide variety of crypto-ransomware.

## 1 INTRODUCTION

Crypto-ransomware is broadly defined as a piece of malicious software which attempts to extort money from a computer user by encrypting the files or documents stored on their machine. [1] Historically crypto-ransomware dates to 1989[2] when Dr. Joseph Popp created the AIDS Trojan. A program mailed out on floppy disk to 7000 subscribers of PC Business World [3] which, when run on an unsuspecting person's PC, would encrypt the user's filenames, making them difficult to identify. Moving forward 28 years we see WannaCry, which affected 300,000 machines across 150 countries including crippling the National Health Service for several days [4].

What happened in twenty years to make this kind of software so much more dangerous? While there are many environmental factors which played a part such as an astronomical increase in the number of machines connected to the internet, an increase in software complexity leading to remotely exploitable security problems, as well the rise in awareness of crypto-currencies like Bitcoin [5]. This paper argues that, even with these things in place, Dr. Popp's Trojan could not have affected the world in the significant way that WannaCry did in 2017.

## 2 ANATOMY OF SUCCESSFUL CRYPTO-RANSOMWARE

### 2.1 Four Necessary Parts

*2.1.1 Agent.* In our model we define this as the active part of the system. It is responsible for encrypting the target machine, contacting any external systems it on which it depends (such as a command and control system), directing victims to provide payment and finally decrypting the target machine when payment is supplied. Malware comes in many different forms and employs various attack vectors. Dr. Popp's earliest attempt was a simple trojan where the user had to be fooled into executing the program on their machine. In this respect modern approaches like Cryptolocker (2013) operate in much the same way. Instead of attempting to trick a target into inserting a disk they are instead tricked into executing an email attachment. [5]. In this sense Dr. Popp's approach is still valid today. This isn't the only attack vector utilized by successful crypto-ransomware agents. The Internet-crippling WannaCry was spread via a known flaw in Microsoft's SMB file sharing protocol [6]. This caused WannaCry to operate more like a worm than Cryptolocker, whose behavior was closer to that of a phishing attack.

In addition to the clear point that without some kind of agent we would have no attacker to defeat, it needs to be stressed that this software cannot be entirely self-contained if it is to be successful. Any software which contains all the necessary information to both encrypt and decrypt the target machine is vulnerable to analysis. Scraper, TeslaCrypt and Radamant are examples of ransomware which failed because researchers were able to examine the program and find a weakness in either the encryption system or the keys themselves [7][8][9]. The need to separate some portion of the encryption/decryption system into component parts was recognized as early as 1996 when researchers Moti Yung and Adam Young published analysis of Dr. Popp's trojan. They noted that it suffered from its use of a symmetric key cipher. Furthermore they suggested that such attacks could be drastically improved by using an asymmetric cipher like RSA [10]. That said, even a system with a single asymmetric key pair or that programmatically generates asymmetric keys would also be quickly defeated. As the agent is also responsible for decrypting the target machine, then either the key or the key-generation algorithm must be stored in the agent itself and could eventually be recovered through analysis. To "succeed" in the same way that WannaCry did, what is needed is a system which produces unique keys for every attack without depending on the agent's own code. AIDS took less than two months to crack. Cryptolocker existed for almost a year and was only defeated due to the efforts of Operation Tovar by recovering the master keys from the Cryptolocker distribution servers. During that period, it took in an estimated $27 million dollars [5].

*2.1.2 Command and Control Service (CNC).* We have already established that the agent requires a source of cryptographic keys which cannot be discovered through code analysis. This is, in effect, what the CNC does. In the case of Cryptolocker the agent makes a request for a cryptographic key pair from the CNC server. The server generates the pair but sends only one to the agent. The agent then begins encrypting the system by first generating a symmetric key cypher for each file then encrypting each file with its own symmetric key. Finally, it encrypts the symmetric key with the asymmetric key provided by the CNC server. Once payment is recognized, the agent asks the CNC server to release the other half of the key pair, stores a copy on the target's desktop and then attends to the tasks of decrypting the target's files. Without the CNC server we have no way to encrypt our files, so success is dependent on our agent being able to find it, but its location on the Internet cannot be static, because that would make it vulnerable to IP or DNS blocking.

*2.1.3 Anonymous Payment Service (APS).* Many sources consider the rise of crypto-currencies key in the development of crypto-ransomware [5]. Whether this a true cause-and-effect relationship or not, anyone who is taking funds illegally would need to protect their identity. Prior to the rise of bitcoin, earlier ransomware like TROJ_CRYZIP.A required payment in E-gold, a now defunct e-currency which, due to the privacy it afforded users, was a favorite for online criminals [11].

*2.1.4 Obfuscated Command Channel (OCC).* Once we have a way of encrypting the machine, a separate source of cryptographic keys and a way to secure payment without getting arrested or our assets seized, all our agent requires to be successful is a method to get those keys and start doing the work. As noted in section 2.1.2, the CNC server needs to avoid IP or domain blocking. Additionally, section 2.1.1 implies that we expect our agent to be examined. Therefore, we may conclude that successful crypto-ransomware can't employ any simple pool of DNS or IP addresses either. We need something much more complex. It is worth noting that our system does not require these transactions to be encrypted. Instead we need them resistant to an analyst predicting exactly where on the Internet our CNC server is going to be. To this end there are two major approaches employed by crypto-ransomware: Tor and Domain Generation Algorithms.

Tor is a network designed to hide the location of an electronic transaction. It achieves this using onion routing, effectively passing your traffic through many other computers, each one only having limited information about its ultimate destination [12]. While Tor may seem like a perfect way to implement an OCC, it may be surprising that it's employed by so few ransomware agents and in fact was entirely unknown until 2014. At that point the well-known blog Malware Don't Need Coffee published an advertisement from the Darkweb[13]. This posting touted a piece of ransomware, which became known as Citroni by researchers of its features the most notable was the use of Tor for CNC communications. Outside of Russia there are few known uses of Citroni and even within Russia it is not known to have generated much income. There is good reason for this. While Tor offers a well-understood way to keep the location of the CNC service secret it has some drawbacks. Nodes can be compromised [14] and having node control or even network control of a node can lead to deanonymization [15]. As well, traffic can be at least partially blocked thus making it an unreliable channel. Ultimately it is an infrastructure heavy solution. Tor requires thousands of nodes to be in operation to offer a reasonable guarantee of privacy.

Most successful crypto-ransomware uses a much simpler approach: DGA. The two most well-known examples of crypto-ransomware: WannaCry and Cryptolocker both employ this technique. Domain Generation Algorithms are simply methods by which an agent and a CNC server can "find" each other by co-opting the Domain Name System (DNS). To do so the agent and server, like spies in a TV drama, need to merely agree to "meet" on the Internet at an appointed time and place (DNS address). This is done by an algorithm which takes as input the date (and possibly the time) and returns a DNS address. For example, if the date was 05/06/2012 our algorithm could direct us to contact FIVESIXTWENTYTWELVE.COM. However, computers, having a distinct speed advantage over your average TV spy, doesn't even have to be that precise. Our software agent could conceivably try a hundred or more domain names without breaking a sweat. This, as we discuss later is a key component of what makes DGA successful even when the agent is under analysis, but for the sake of this example our algorithm could direct our agent to attempt to contact to the CNC server at FIVESIXTWENTYTWELVE1.COM, If it doesn't find the server there it's algorithm might tell it to continue to try FIVESIXTWENTYTWELVE2.COM then FIVESIXTWENTYTWELVE3.COM and so on. All that's required is for our server to be waiting at one of those domains at the given day and time. This allows for the server to constantly change its IP address and domain name to avoid counter-measures.
.

**2.3 Countermeasures**

*2.3.1 Defense against the agent:* The methods for foiling the agent are heavily dependent on the attack vector employed. For example, agents that attack the local system can be handled by anti-virus and anti-malware applications. Vendors of this kind of software are constantly attempting to detect new and emergent threats to people's systems. Countermeasure falls into two broad categories: signature-based or behavioural. Signature based systems take the bytes that make up the agent and by some mechanism (i.e. a hashing function) create a short relatively unique set of bytes which represent some part of the file's binary image. This is called a signature. Anti-virus software keeps a list of signatures matching dangerous software and scans incoming information like email attachments or even your entire hard drive. When a file matches the signature, the software takes remedial action because of this signature-based system are very accurate. However, since we need an example of the agent in order to generate a signature our system can only react to threats that we already know about. As WannaCry illustrates, to us that signature might arrive too late. Furthermore, signatures may also be fooled by padding – adding extra information to a file or encryption. [16] To better protect users anti-virus software usually employs behavioural methods as well. Instead of looking at what bytes make up the agent, the software now examines what the agent does (e.g. re-write a file with a higher amount of entropy). This approach is employed specifically to combat crypto-ransomware by software like Sophos's Intercept X. That said, the inexact nature of behavioural analysis (zipping files and legitimate encryption also increase entropy) causes these kinds of countermeasure have a much higher false positive rate.

Network-based attack vectors can be stopped by ordinary networking tools like firewalls – blocking port 445 at the edge of your network was a common suggestion to prevent WannaCry [17]. Signature and behavioural analysis can be used here as well. Fox-it, one of the groups involved in the takedown of the Cryptolocker servers has produced scripts for the popular network monitoring application Bro to check the entropy of SMB transfers [18]. Effectively telling us if a file being written has been encrypted. Since encryption must also increase entropy any SMB write requests of this kind would be suspect. With the exception of entropy checking, these methods are also reactive to the particular attack vector. And since many common activities legitimately increase entropy, doing entropy checks might only be useful to warn the user to act, at which point some of our files have already been encrypted.

*2.3.2 Defense against command and control service:* Disrupting the CNC service is nothing new. The so-called WannaCry killswitch [19] is effectively a built-in method of keeping the agent from contacting the CNC service. Check Point, a security hardware/software vendor, introduced a "sinkhole" method into their managed security service "ThreatCloud," which blocks CNC traffic [20]. However, both of these methods are reactive and carry the same concerns that signature based methods do: We may not understand the threat until it is too late.

*2.3.3 Defense against secure payment channel:* As Bitcoin and other cryptocurrencies are probably not going away any time soon, the defenders have few options here. The only well-known examples of a attacks against a payment channel involve denial of payment. For example during the Petya/Not-Petya attack, the agent would provide the target with an email address where a target could send their bitcoin wallet ID as proof of payment. This email address happened to be hosted by Posteo, an email provider based in Berlin, Germany. When the administrators realized their system was being used in a malware attack they shut it down [22]. Unfortunately while keeping the criminals from getting paid this did little to protect the people whose files were now trapped on their machines in an encrypted form.

*2.3.2 Defense against obfuscated command channel:* Defense against Tor is a well-understood idea. Many firewalls have options for blocking it (or attempting to block it) [23]. DGA is another matter. Since DNS is a service so integral to the operation of Internet applications it is likely impossible to shut down or limit and while our trivial example in section 2.14 might seem like it's plausible to simply restrict every address the DGA algorithm can produce. The reality is not so simple. The Conflicker.A and Conflicker.B worms, which popularized this technique before ransomware got a hold of it used an algorithm which generated 250 different domains a day. The Conflicker.C worm would generate 50,000[23]. Attempting to pre-generate every address would be difficult. However, since a server or individual registering 50,000 domains might be a bit conspicuous the Conflicker.C's CNC server only registers a randomly selected 1% of these potential addresses, which is easily hidden in the noise of DNS registries. This doesn't affect the success rate of the agent though. It only needs to remain undetected while it waits to find the CNC server. It's also worth pointing out that even generating the list of domain names to block each day still results in another reactive system. In the case of a wide-spread attack this kind of information would arrive well after the damage was done.

However, it is here we suggest a new defensive method. To avoid DNS- and IP-blocking defences. The OCC absolutely depends on the *timely setup of a domain name – in the order of hours.* As we will illustrate later on, there are so few services which require access to such recently minted domain names that we may be able to defend against virtually all crypto-ransomware with a DGA-based OCC simply by denying use of DNS addresses that are newly created.

## 3   Methods for delaying DNS resolution.

The DNS standard is like most of the key parts of the internet: Very old. It was proposed by Paul Mockapetris in 1983 and was designed to solve a very specific problem: Maintaining the increasingly complex HOSTS.TXT files [24]. Thus, it did not include any facility for determining exactly when a record was created. There are a few online search engines like SecurityTrails [25]. However, these were designed to retrieve historical information. They are not expected to be accurate to a matter of hours.

One source that should be accurate is the DNS record itself. Domains that have a different cutover point from the parent domain will have a SOA record. It is recommended by RFC1912 that the serial number field contains a "last changed time" in the ISO 8601 format [26]. While this is a clever idea it depends on too many assumptions: That every domain is going to have its own SOA record and that they follow the standard.

However, there are now several services which, on a subscription basis, provide downloadable CSV files containing recent domain registrations or even access to an interactive API [27]. Tools like these could be easily scripted into a form which could be consumed by IPS tool like Snort or fire off an alert via an IDS tool like Bro.

### 3.3   FALSE POSITIVES AND POTENTIAL HARM AND AVOIDANCE TECHNIQUES.

Our team has been using Bro 2.5 as a logging and early-warning system for about a year now. After recording a sample of DNS traffic for several months my team was unable to find a single instance of a DNS request which was registered only 24 hours earlier. If we configured our system to automatically block DNS requests to these domains, then even if a false positive were to occur it would self-resolve. Triggering an alert in that period would give our network operations team ample time to find, isolate and respond to this condition. Especially given the apparent rarity of the event.

What other kinds of OCC are there to be exploited? Clearly any service which can be expected to exist on an Internet-connected machine is a good candidate. However, few are going to be as ubiquitous as DNS. That said, since successful ransomware also requires an anonymous payment system (APS) it's possible that these two could be combine. There already exist facilities for embedding messages in the bitcoin blockchain such as Eternity Wall (https://eternitywall.it/). Similarly it's possible for the agent to be its own distributed service. Young and Yung outline a virus utilizing this approach in their aforementioned paper [10].

## 4 CONCLUSIONS

In summary, we have demonstrated that successful crypto-ransomware requires an Obfuscated Command Channel, which in successful crypto-ransomware depends on timely updates of their DNS records. While there are many counter-measures for this kind of attack, few are both precise and proactive. We have illustrated that by carefully denying access to domains that have been newly created we can eliminate this class of malware from affecting our systems.

Created August 2017; revised August 2018;